\documentclass{article}

\usepackage{moreverb,url}
\usepackage[colorlinks,bookmarksopen,bookmarksnumbered,citecolor=red,urlcolor=red]{hyperref}
\usepackage{physics}
\usepackage[sort&compress]{natbib}
\usepackage{float}
\RequirePackage{graphicx}
\RequirePackage{latexsym,ifthen,rotating,calc,textcase,booktabs,color,endnotes}
\RequirePackage{amsfonts,amssymb,amsbsy,amsmath,amsthm}

\newcommand{\pd}[2]{\frac{\partial #1}{\partial #2}} 
\newcommand{\bfW}{{\bf W}}	
\newcommand{\bfX}{{\bf X}}
\newcommand{\bfY}{{\bf Y}}
\newcommand{\bfZ}{{\bf Z}}

\newcommand{\bfR}{{\bf R}}
\newcommand{\bfS}{{\bf S}}

\newcommand{\bfU}{{\bf U}}

\newcommand{\bfM}{{\bf M}}
\newcommand{\bfy}{{\bf y}}
\newcommand{\bfx}{{\bf x}}

\newcommand{\bfz}{{\bf z}}

\newcommand{\bfs}{{\bf s}}

\newcommand{\alphabf}{\boldsymbol{\alpha}}
\newcommand{\betabf}{\boldsymbol{\beta}}

\newcommand{\thetabf}{\boldsymbol{\theta}}
\newcommand{\deltabf}{\boldsymbol{\delta}}

\newcommand{\gammabf}{\boldsymbol{\gamma}}

\newcommand{\Thetabf}{\boldsymbol{\Theta}}

\newcommand\BibTeX{{\rmfamily B\kern-.05em \textsc{i\kern-.025em b}\kern-.08em
T\kern-.1667em\lower.7ex\hbox{E}\kern-.125emX}}

\begin{document}

\title{Monitoring of process and risk-adjusted medical outcomes using a multi-stage MEWMA chart}

\author{Doaa Ayad, Nokuthaba Sibanda \\
School of Mathematics and Statistics\\
Victoria University of Wellington\\
New Zealand\\
nokuthaba.sibanda@vuw.ac.nz}
\date{28 May 2020}
\maketitle

\begin{abstract}
Most statistical process control programmes in healthcare focus on surveillance of outcomes at the final stage of a procedure, such as mortality or failure rates. Such an approach ignores the multi-stage nature of these procedures, in which a patient progresses through several stages prior to the final stage. In this paper, we develop a multi-stage control chart based on a multivariate exponentially weighted moving average (EWMA) test statistic derived from score equations. This allows simultaneous monitoring of all intermediate and final stage outcomes of a healthcare process, with adjustment for underlying patient risk factors and dependence between outcome variables. Use of the EWMA test statistics allows quick detection of small gradual changes in any part of the process. Three advantages of the approach are: better understanding of how outcomes at different stages relate to each other, explicit monitoring of upstream stage outcomes may help curtail trends that lead to poorer end-stage outcomes and understanding the impact of each stage can help determine the most effective allocation of quality improvement resources. Simulations are performed to test the control charts under various types of hypothesised shifts, and the results are summarised using out-of-control average run lengths. 

\end{abstract}

%\keywords{multi-stage control chart, healthcare statistical process control, EWMA, profile monitoring}

\section{Introduction}

\label{sec1}

Interest in monitoring and continuous improvement of medical outcomes has grown steadily over the last three decades. Results of monitoring exercises are used by regulatory organizations, hospital administrators, medical practitioners and other stakeholders to detect changes in performance and to compare practitioners or  different hospitals.  Current methods of monitoring medical outcomes focus on final outcomes of a given procedure, such as surgical failure rate.  The methods used are based on those developed for monitoring in industrial contexts \citep{montg2009}.  To take into account heterogeneity of expected outcomes among patients, risk-adjusted monitoring was proposed \citep{lovegrove1997, poloniecki1998, steiner2000}. In risk-adjusted monitoring, the expected outcome is not uniform among patients, but rather depends on key risk factors of the patient. Some recent developments include simultaneous monitoring of multiple outcomes \citep{waterhouse2012} and monitoring that allows for underlying changes in performance over time \citep{steiner2014}.  

In addition to patient-specific risks, there is heterogeneity among practitioners in the decisions they make and the processes they follow during the procedure. This issue has not received much attention in the literature.  Furthermore, the final outcome of a procedure may be related to outcomes of preceding stages within the procedure.  To address both issues, \cite{sibanda2016} proposed monitoring outcomes of intermediate stages in addition to the final outcome, adjusting for heterogeneity in patient risk and processes used by practitioners.  A model for the procedure in its entirety was developed incorporating patient risks, processes and outcomes at all stages. A test statistic based on the ratio of observed ($O$) to expected ($E$) number of failures in consecutive patient subgroups was used for each outcome.  While the $O/E$ test statistic has a clear clinical interpretation, it is slow in detecting small shifts and is unsuitable for rare outcomes. In this article we propose monitoring  the regression coefficients of the model that represents the relationships among the outcomes, processes and patient risk factors relevant to the procedure. To ensure quick detection of small gradual shifts we use a sequential monitoring approach where the chart statistic is updated at each successive observation. 

In the statistics and econometrics literature, considerable attention has been paid to problems of monitoring constancy of parameters in statistical models. \cite{nyblom1989} derived a Lagrange multiplier (LM) test based on likelihood score equations for a time-series model and showed this to be the locally most powerful test for the alternative that the parameters follow a martingale process. \cite{hansen1992} extended the LM test to linear regression models, and \cite{hjort2002} suggested a general class of likelihood score-based structural change tests. In this article, we develop a monitoring procedure for model coefficients based on likelihood score equations.

In the statistical process control literature, profile monitoring is used for model based processes where the quality of the process or a stage of the process is represented by a functional relationship between a response variable and one or more explanatory variables. \cite{noorossana2011} describe profile monitoring in a number of settings. The main objective of profile monitoring is to check the stability of a functional relationship over time. \cite{mandel1969} and \cite{hawkins1993} give examples of model based profile monitoring procedures. In these references, the response variable was assumed to follow a Normal distribution and regression coefficients were estimated using ordinary least squares (OLS). \cite{skinner2003} and \cite{jeark2003} incorporated use of the generalized linear model in profile monitoring of Poisson and Gamma distributed response variables. \cite{jeark2005} and \cite{Shang2013} developed a procedure based on monitoring deviance residuals using Shewhart charts for a multistage process with non-Gaussian response variables. \cite{Shang2013} pointed out that an exponentially weighted moving average (EWMA) scheme would be required to detect small shifts in the multistage process. \cite{yeh2009} developed a profile monitoring procedure using logistic regression for a single binary response variable. In this paper, we develop a multivariate EWMA (MEWMA) procedure based on likelihood score equations for simultaneous monitoring of the vector of model coefficients of a multistage process. In this way we monitor stability of the functional relationship between multiple response variables and associated predictors.

There are a number of issues to consider when monitoring a multivariate vector of parameters in a  model. For example, monitoring  model parameters is more complex than the traditional SPC approach of monitoring the response variables themselves. Also, a medical procedure can have  categorical outcome variables.  Our approach is flexible in that it can handle such  processes, since the likelihood function and score equations can be obtained in a straightforward manner once the assumed distributions of the response variables are known.
In the next section, we describe a motivating example and represent the process using a graphical model presentation.  Then in Section \ref{sec.mewma}, we present the general formulation of our proposed monitoring procedure, together with derivation of the associated chart statistic. Performance of the proposed chart is assessed using simulation studies in Section \ref{sec.simulation} and the results are discussed in Section \ref{sec.discussion}.

\section{Motivating example}
\label{sec.example}

The motivating example is the delivery process in a maternity unit. The example  is described in detail by  \cite{sibanda2016} and is revisited briefly here.  The medical procedure of interest is infant delivery in the labour ward of a maternity unit.  Four maternal outcomes, represented by $Y =(Y_1, Y_2, Y_3, Y_4)'$, are identified.  These outcomes are temporally ordered as they occur at various ordered stages of the process.  Temporal ordering allows us to represent correlations among the outcome variables  using a  graphical model framework. In addition to being influenced by outcomes at upstream  stages, each $Y_v, v=1,\ldots,4$ depends on a number of process factors, $X$, and patient risks, $Z$ that can also be included in the graphical model.  The graphical representation of a procedure extends the literature on the regression adjustment approach and multistage control charts developed in industrial applications \citep{hawkins1991, hawkins1993, zantek2002}. The graphical representation of the delivery procedure is shown in Figure \ref{fig.stages}.

\subsection{Model specification for the delivery procedure}
\begin{figure}[H]
\centering
\includegraphics[scale=0.6]{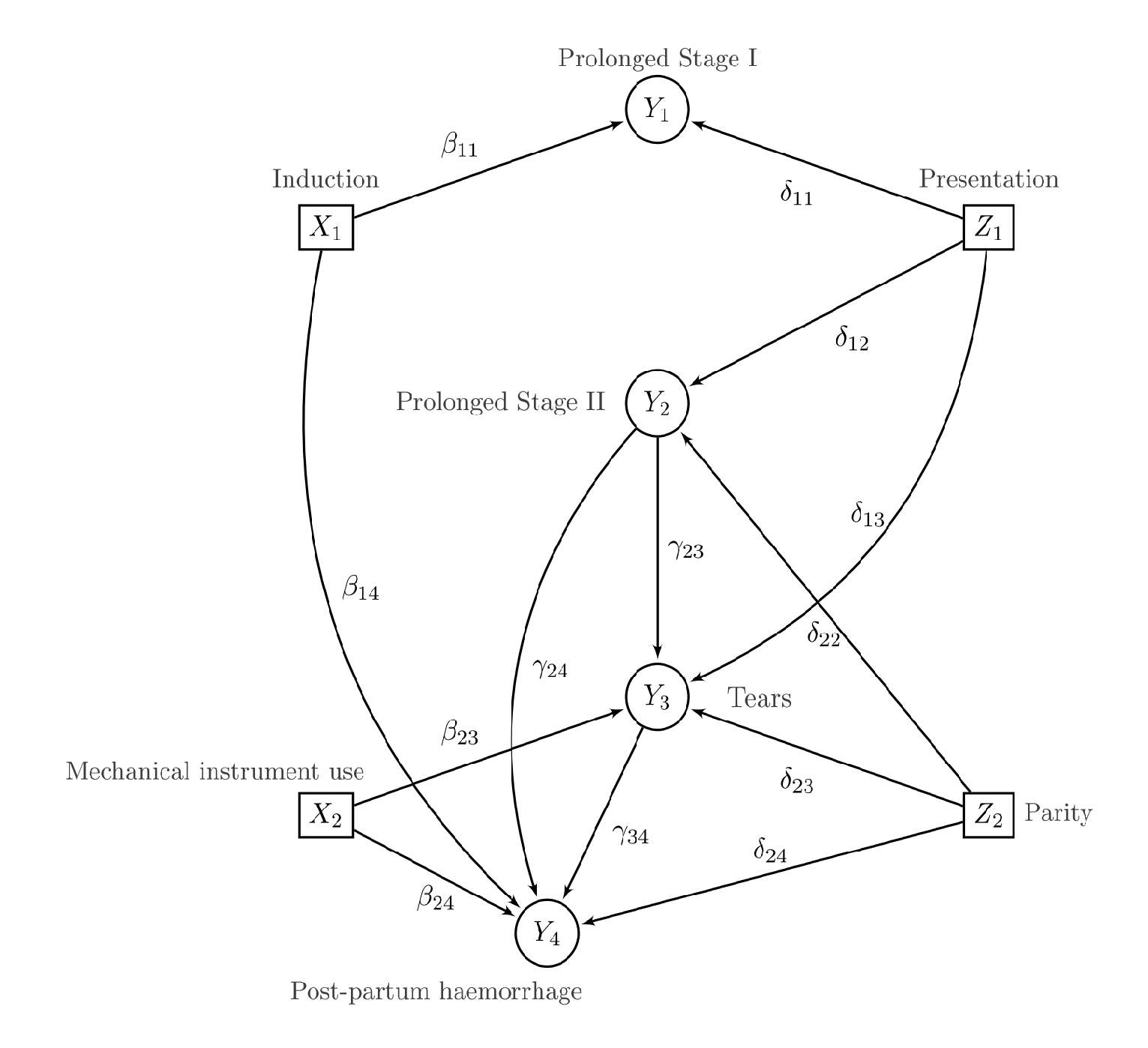}
\caption{\label{fig.stages} Graphical representation showing the model structure for a multi-stage procedure in a maternity unit. Edge directions indicate the direction of relationships and edge labels are the coefficients for the model in equation (\ref{eq:processspec}). The outcome variables are: $Y_1$=Prolonged Stage I labour, $Y_2$=Prolonged Stage II labour, $Y_3$=3rd or 4th degree tears and $Y_4$=Post-partum haemorrhage. The process variables are: $X_1$=Induction of labour and $X_2$=Mechanical instrument use. The risk factors are: $Z_1$=Posterior or transverse presentation and $Z_2$=Parity.}
\end{figure}

There are three major stages in the delivery process with a total of four maternal outcomes defined as: 
{\small
\begin{eqnarray*}
Y_1 &=& \begin{cases} 1 & \text{ if no prior births and $L_1 > 18$ hours, or if $\ge 1$ prior births and $L_1 > 12$ hours} \\
							0 & \text{ otherwise} \end{cases} \\
Y_2 &=& \begin{cases} 1 & \text{ if no prior births and $L_2 > 2$ hours, or if $\ge 1$ prior births and $L_2 > 1$ hour} \\
							0 & \text{ otherwise} \end{cases} \\
Y_3 &=& \begin{cases} 1 & \text{ if a 3rd or 4th degree tear occurs} \\
							0 & \text{ otherwise} \end{cases} \\
%Y_4 &=& \begin{cases} 1 & \text{ if 5-minute Apgar score $<7$} \\
							%0 & \text{ otherwise} \end{cases} \\
Y_4 &=& \begin{cases} 1 & \text{ if maternal blood loss $>500ml$ with no Cesarean, or $>1000ml$ with a Cesarean} \\
							0 & \text{ otherwise}, \end{cases} 													
	\label{eq:yvardefine}
\end{eqnarray*}
}
%The outcome variables are analysed here in binary form to capture only the variation relevant in determining rates of `adverse' outcomes.  Since  the aim in quality improvement is to reduce rates of adverse outcomes, discretizing the variables in this way is more appropriate than analysing the variables in their raw form.
%In practice, interest is in whether or not labour is pro-longed and $Y_1$ and $Y_2$ are usually transformed to binary variables.   
Patients who do not undergo labour due to an elective or emergency Cesarean section have been omitted from this study since the only outcome observed for them  is $Y_4$.   Each outcome variable is influenced by one or more process variables given by
\begin{eqnarray*}
X_1 &=& \begin{cases} 1 & \text{ if labour induced }\\
							0 & \text{ otherwise} \end{cases} \\
X_2 &=& \begin{cases} 1 & \text{ if mechanical instruments used during the second stage}\\
							0 & \text{ otherwise} \end{cases} \\
\label{eq:xvardefine}
\end{eqnarray*}							
					
and risk factors given by	
\begin{eqnarray*}							
Z_1 &=& \begin{cases} 1 & \text{ if presentation is posterior or transverse } \\
							0 & \text{ otherwise} \end{cases} \\
Z_2 &=& \begin{cases} 1 & \text{ if it is the mother's first birth (Parity=0)} \\
							0 & \text{ otherwise (Parity $\ge 1$).} \end{cases} 
%Z_3 &=&  \begin{cases} 1 & \text{ if gestation $< 37$ weeks (pre-term birth)} \\
							%0 & \text{ otherwise} \end{cases} \\												
\label{eq:zvardefine}
\end{eqnarray*}			
A combination of literature \citep{thorngren2001, cameron2006, rouse2009,hirayama2012} and empirical evidence was  used to determine the structure of the graphical presentation as shown in Figure \ref{fig.stages}.

The graphical model in Figure \ref{fig.stages} can be viewed as a Bayesian network \citep{pearl1986, lauritzen1988}. Note that for clarity of presentation we have deviated from standard graphical model presentation by not including separate edges for model coefficients. A Bayesian network is a directed acyclic graph (DAG) with node set $V$ representing a set of random variables, $\bfY=\{Y_{v\in V}\}$ having a joint probability distribution function that can be written as
\begin{equation}
	P(\bfY)=\prod_{v \in V}P(Y_v|Y_{\textrm{pa}(v)}).
	\label{eq:bayesnet}
\end{equation}
The term $\textrm{pa}(v)$ represents the set of parent nodes of the node $v$. The power of a DAG representation is that once the structure is known, the joint probability distribution of $\bfY$ can be written in the form of equation (\ref{eq:bayesnet}) using the conditional independence axioms introduced by \cite{dawid1979}. In equation (\ref{eq:bayesnet}), each node is conditionally independent  of all other nodes, given its parent nodes. Based on the graphical structure in Figure \ref{fig.stages} we can write:

\vspace*{-1.5em}

\begin{align}
& P(Y_{1},Y_{2},Y_{3},Y_{4}|\bfX,\bfZ,\Thetabf) \nonumber \\
&= P(Y_{4}|Y_{2},Y_{3},\bfX,\bfZ,\thetabf_4)P(Y_{3}|Y_2,\bfX,\bfZ, \thetabf_3)P(Y_{2}|\bfX,\bfZ, \thetabf_2)P(Y_{1}|\bfX,\bfZ, \thetabf_1) \nonumber
	\label{eq:cond}
\end{align}
where $\Thetabf = (\thetabf_1, \thetabf_2, \thetabf_3, \thetabf_4)'$ is a vector of parameters characterizing the relationship between variables.

\subsection{Process model and monitoring}
For each patient, $t$, we observe the outcome vector $\bfy_t$, the process vector, $\bfx_t$ and the risk-factors $\bfz_t$. Using generalized linear model (GLM) notation we write

\begin{equation}
g_v(\mu_{vt})  = \alpha_v + \bfx_{vt}'\betabf_v + \bfy_{pa(v)t}'\gammabf_v + \bfz'_{vt}\deltabf_v
\label{eq:processgen}
\end{equation}
where $g_v()$ is an appropriate link function and $\mu_{vt} = E(Y_{vt})$,  for $v=1,2,3,4$ and $t=1,2,\ldots, n$. In general, the process model may comprise a mix of continuous, discrete, ordinal or nominal variables. In such cases, it is possible to have a different link function for each outcome variable $Y_v$.  In our case all outcome variables are binary in nature and here we choose to use the logit link for all $Y_v, v=1,\ldots,4$.  We therefore have:

\begin{equation}
g_v(\mu_{vt}) = \log\left(\frac{\mu_{vt}}{1-\mu_{vt}}\right) =\alpha_v + \bfx_{vt}'\betabf_v + \bfy_{pa(v)t}'\gammabf_v + \bfz'_{vt}\deltabf_v,
\label{eq:processspec}
\end{equation}
for all $v \in \{1,2,3,4\}$ and $\mu_{vt} \in (0,1)$.  Our proposed procedure is based on monitoring the model coefficient vector $\Thetabf = (\thetabf_1, \thetabf_2, \thetabf_3, \thetabf_4)'$ where $\thetabf_v= (\alphabf_v, \betabf_v, \gammabf_v, \deltabf_v)'$ for all $v \in \{1,2,3,4\}$.  

\section{Score based Multivariate EWMA chart \label{sec.mewma}}
The score-based multivariate EWMA (MEWMA) chart is the monitoring scheme we introduce to overcome the problems associated with rare outcomes and detection of small shift  \citep{sibanda2016}. The MEWMA \citep{lowry1992}, accumulates information from past observations for quick detection of small  shifts. Given a vector, $\bfM_t = (M_{t1}, \ldots, M_{tV})'$, of $V$  quality indicators for patient $t$, the MEWMA chart is based on the statistic 
$$T^2_t =  \bfW'_t \Sigma_{\bfW_t}^{-1} \bfW_t,$$
where the MEWMA vector $\bfW_t $ is calculated using 
$$\bfW_t = \bfR\bfM_t + (\mathbb{I}-\bfR)\bfW_{t-1} =\sum_{i=1}^t \bfR(\mathbb{I}-\bfR)^{t-i}\bfM_i,$$
and $\Sigma_{\bfW_t}$ is the covariance matrix for $\bfW_t$ given by $$\Sigma_{\bfW_t}=\sum_{i=1}^t \bfR(\mathbb{I}-\bfR)^{t-i}\Sigma_\bfM (\mathbb{I}-\bfR)^{t-i}\bfR.$$
$\bfR$ is a $V \times V$ matrix with $j^{th}$ diagonal element $r_j \in (0, 1]$ for $j=1,2,\ldots,V$, and zero elements on the off-diagonal. $\mathbb{I}$ is the identity matrix and $\Sigma_\bfM$ is the covariance matrix for $\bfM_t$.   The parameter $r_j$ is called the smoothing parameter, and it determines the relative weight of current and past observations for the $j^{th}$ quality characteristic. If there is no \emph{a priori} reason to use different weights for the $V$ quality characteristics, the diagonal elements of $R$ are all set equal to a constant $r$.  When $r_1 = r_2 = \ldots = r_V=r$, \cite{montg2009} shows that the covariance matrix simplifies to $$\Sigma_{\bfW_t}=\frac{r[1-(1-r)^{2t}}{2-r}\Sigma_{\bfM}.$$  The MEWMA chart signals at the first observation at which $T^2_t > h$, where $h>0$ is a control limit chosen to achieve a specified in-control average run length (ARL).

In our case, the multivariate quality indicator of interest is the vector of regression coefficients for the functional relationships between the response variables and the process variables and risk factors. To detect a change in the functional relationship, we propose monitoring the score vector of the regression coefficients.  For a graphical model with $V$ response variables $Y_1, Y_2, \ldots, Y_V$, process variables $\bfX$, patient-risk variables $\bfZ$ and parameter vector $\Thetabf$ of length $p$, the likelihood function is given by

\begin{equation*}
L(\Thetabf|\bfy,\bfx,\bfz) = \prod_{v=1}^V f_v(\bfy_{v}|\bfy_{pa(v)},\bfx,\bfz, \thetabf_v),
\label{eq:}
\end{equation*}
where $\bfy_v = (y_{v1}, y_{v2},\ldots,y_{vn})'$ is a vector of observed values of $Y_v$ for $n$ patients and $f_v()$ is the density or probability distribution function of $Y_v$. The log-likelihood function is given by

\begin{equation}
\ell(\Thetabf|\bfy,\bfx,\bfz) = \sum_{v=1}^V\sum_{t=1}^n \log f(y_{vt}|\bfy_{pa(v)t},\bfx_t,\bfz_t, \Thetabf).
\label{eq:loglik}
\end{equation}
Let $\bfs(\bfy,\Thetabf|\bfx,\bfz)$  be  a $p-$length vector of first  derivatives of $\ell(\Thetabf|\bfy,\bfx,\bfz)$, $$\bfs(\bfy,\Thetabf|\bfx,\bfz)=\pd{\ell(\Thetabf|\bfy,\bfx,\bfz)}{\theta_j}.$$ 

Then $ \dfrac{1}{\sqrt n} \bfs(\bfy,\Thetabf|\bfx,\bfz) $ has an asymptotic $p-$ variate  normal distribution with mean zero and covariance matrix given by the Fisher Information matrix, $I=-E[i(\bfy,\Thetabf|\bfx,\bfz)]$, where $i(\bfy,\Thetabf|\bfx,\bfz)$ is a matrix of second derivatives of $\ell(\Thetabf|\bfy,\bfx,\bfz)$ with respect to $\Thetabf$  \citep{rao1973}.

A common approach for testing the stability of a $p-$dimensional vector of regression coefficients is based on a cumulative sum of standardised scores at observation $t$ of the form 

\begin{equation*}
M_n(t) = I^{-1/2}\frac{1}{\sqrt{n}} \sum_{i\le t} S(\bfy_i,\Thetabf_0|\bfx_i,\bfz_i) \quad \text{for } 1 \le t \le n,
\label{eq:} 
\end{equation*}

where $\Thetabf_0$ is the specified value of $\Thetabf$ under the null hypothesis of no change in the regression coefficients. The components of $M_n(t)$ tend in distribution to $p$ independent Brownian motions under the null hypothesis \citep{hjort2002} and are used for separate tests of the $p$ parameters. When $\Thetabf$ is unknown, a maximum likelihood estimate $\hat\Thetabf$ can be determined from data from a previous time period (Phase 1 data) and the estimated cumulative standardized score becomes

\begin{equation*}
M_n(t) = \hat{I}^{-1/2}\frac{1}{\sqrt{n}} \sum_{i\le t} S(\bfy_i,\hat\Thetabf|\bfx_i,\bfz_i) 
\quad \text{for } 1 \le t \le n.
\label{eq:}
\end{equation*} 
Our approach is based on the cumulative sum of standardised scores, but with different weights given to current and past observations.  This gives a score-based multivariate exponentially weighted moving average (MEWMA) chart. This score-based MEWMA chart gives greater weight to more recent observations and will therefore be quicker at detecting smaller gradual changes in the process parameters. 

When the MEWMA is applied to the score vector $\bfs(\thetabf)$, we calculate 
$$\bfW_t = \bfR\bfS_t + (\mathbb{I}-\bfR)\bfW_{t-1} =\sum_{j=1}^t \bfR(\mathbb{I}-\bfR)^{t-j} \bfS_j $$
at time $t$, for i.i.d $\bfS_t$ with $E(\bfS_t) =\mathbf{0}_p$ and $Var(\bfS_t)=\Sigma_\bfs = -E\left[\pdv{\ell}{\thetabf}{\thetabf'} \right]$ if the process is in control.
This gives $E(\bfW_t) =\mathbf{0}_p$ and $Var(\bfW_t)= \Sigma_{\bfW_t} =\sum_{j=1}^t \bfR(\mathbb{I}-\bfR)^{t-j}\Sigma_\bfS (\mathbb{I}-\bfR)^{t-j}\bfR$.  We would then chart
$$T^2_t =  \bfW'_t \Sigma_{\bfW_t}^{-1}\bfW_t,$$
with control limit $h$ identified through simulation to achieve a desired in-control $ARL$.

\subsection{Score equations and information matrix for the delivery process}
For each $Y_v$, let $\thetabf_v = (\alpha_v, \betabf_v, \gammabf_v, \deltabf_v)'$ be a $p_v$-dimensional vector and $\bfU_v = (\mathbf{1}, X_v, Y_{pa(v)}, Z_v)$, where $p_v$ is the number of regression coefficients for $Y_v$ and $\bfU_v$ is the corresponding design matrix.  For the model in Figure \ref{fig.stages} and equation \ref{eq:processspec}, the log-likelihood function is given by 

\begin{align}
\ell(\Thetabf) &= \sum_{v=1}^V \sum_{t=1}^n \left( y_{vt}\log\left(\frac{p_{vt}}{1-p_{vt}}\right) + \log \left(1- p_{vt}\right)\right) \nonumber \\
&= \sum_{v=1}^V \sum_{t=1}^n  \left( y_{vt}(\thetabf'_v U_{vt}) -\log (1 + \exp(\thetabf'_v U_{vt}))\right),
\label{eq:loglik}
\end{align}
with $p = \sum_v p_v$-dimensional score vector with elements 
\begin{equation}
\bfs(\thetabf_v) = \sum_{t=1}^n u_{vt}\left[y_{vt} - \left(\frac{\exp(\thetabf'_v U_{vt})}{1 +\exp(\thetabf'_v U_{vt}) }\right)\right],
\end{equation}
for $v=1,\ldots,V$. The observed Fisher information matrix is a block diagonal matrix given by
$$I(\boldsymbol{\Theta}) = \begin{bmatrix} I(\boldsymbol{\theta}_1) & {0} & \cdots & {0} \\
		{0} & I(\boldsymbol{\theta}_2) & \cdots & { 0} \\
																	\vdots & \vdots & \ddots & {0} \\
																		{ 0} & { 0 } & \cdots & I(\boldsymbol{\theta}_V)
			\end{bmatrix},$$
		where 
$$I(\thetabf_v) = \sum_{t=1}^n u_{vt}p_{vt}(\thetabf_v)(1-p_{vt}(\thetabf_v)).$$

The block diagonal structure of the information matrix suggests a natural partition of the regression coefficient vector into independent sets that can be used to monitor subsets of the coefficients. 

\section{Simulation study \label{sec.simulation}}

In this section, we present a simulation procedure and results for how the score-based MEWMA chart reacts to different types of shifts or changes at different stages of the multistage process for a large tertiary hospital. We check the efficiency of this score based chart in detecting small and large shifts in the process overall quality due to changes in process variables, risk factors or in upstream stage outcomes.
% Different changes in $\Thetabf_i$; of one or more of $\bfX$, $\bf Y$ or $\bfZ$; on $Y_i$ are investigated.
 The aim of the simulation is to obtain empirical estimates of average run length (ARL) as a measure of chart performance. The upper control limit used in this case is the one obtained when we set the chart to achieve in control ARL of 200 under the assumption that Phase I parameters are known without error. We used data from the Southmead maternity unit in Bristol, updated from that published in a previous paper \citep{sibanda2007} to obtain model parameter estimates and these are shown in
Table \ref{tab:parest}. These model parameter estimates are used as the basis for the simulation procedure for testing the proposed control chart.  

\begin{table}[h]
\small\sf\centering
  \caption{Parameter values for the logistic regression models represented in Fig \ref{fig.stages} and equation (\ref{eq:processspec})\label{tab:parest}}
\begin{tabular}{@{}cc@{}} 
\hline
Parameter & Value \\
\hline 
    $  \beta_{11}$ & -1.724 \\
    $  \delta_{11}$ & 0.730 \\
    $  \delta_{12}$ & 1.682 \\
    $  \delta_{22}$ & 1.262 \\
    $  \beta_{23}$ &  0.597\\
    $  \gamma_{23}$ & 0.342  \\
    $  \delta_{13}$ & 0.467 \\
    $  \delta_{23}$ & 0.758 \\
    $  \beta_{14}$ & 0.316 \\
    $  \beta_{24}$ &  1.140\\
    $  \gamma_{24}$ & 0.482 \\
    $  \gamma_{34}$ & 1.267 \\
    $  \delta{24}$ & 0.374 \\
		\hline
\end{tabular}
\end{table}

There is an average of about 6400 deliveries a year at Southmead Hospital, which gives an average of 533 deliveries a month. Hence, for each hypothesised shift, a single data set of size 500 patients was generated for model development to correspond to the size of dataset available for monitoring at monthly intervals. This size of 500 fulfills the asymptotic assumption of the score-based MEWMA chart. We consider four types of shifts.
\begin{enumerate}
	\item An additive shift in a given regression coefficient(s): $ \mu_{vt_1} =  g^{-1}(\eta_{vt_{-j}}+u_{vjt}(\theta_{vj}+c\theta_{vj}))$, where $\eta_{vt_{-j}}$ is the linear predictor excluding the $j^{th}$ term. 
	\item Simultaneous additive shifts in pairs of regression coefficients.
	\item An additive shift in the mean response: $\mu_{vt_1}=\mu_{vt_0}(1+c)=g^{-1}(\thetabf'_vU_{vt})(1+c)$. Values of $c$ were restricted to where $\mu_{vt_1} \in (0,1)$.
	\item A more appropriate approach for binary outcomes is to consider shifts in the odds ratio \citep{steiner2000}.  Under current conditions, we assume an odds ratio, $R_{v0}$ of 1.  If a shift occurs in one of the outcome variables the odds ratio is $R_{v1}=c$ so that 
	$$R_{v1}=\frac{\mu_{vt_1}/(1-\mu_{vt_1})}{\mu_{vt_0}/(1-\mu_{vt_0})} =c.$$
	This means $$\mu_{vt_1}=\frac{c\mu_{vt_0}}{1-\mu_{vt_0}+c\mu_{vt_0}} .$$
	
	%\item A multiplicative shift in the mean of an upstream outcome variable that affects a downstream outcome: 
\end{enumerate} 

For each shift type, a range of values of $c$ are used and for each value of $c$, the chart is constructed 5,000 times and the run length recorded for each run. The arithmetic mean of the run lengths is used to estimate the out-of-control ARL at which the score-based MEWMA chart is likely to detect the proposed shift.

First, we study performance of the chart when the shift in the functional relationship occurs in $\beta_{lv}$, the effect of  process variable $X_l$ on $Y_v$ when $E(X_l)$ remains unchanged. These kinds of changes are likely to occur with changes in staff or equipment. For illustration, changes of various sizes are introduced to $\beta_{23}$ and $\beta_{24}$, the effects of $X_2$ (use of mechanical instruments) on $Y_3$ (occurrence of 3rd or 4th degree tear) and on $Y_4$ (occurrence of postpartum haemorrhage), respectively. The empirical out-of-control ARLs under various shifts are shown in Table \ref{tab:ARL1}, indicating that  detecting a change in the functional relationship of the process depends on the size of change and the size of the parameter affected. In this case, $\beta_{23} < \beta_{24}$, and when an equivalent change is applied to both parameters, with all other parameters kept constant, detection of changes occurs more quickly for a shift in $\beta_{24}$ than it does for a shift in $\beta_{23}$.

	\begin{table}[h]
	\small\sf\centering
		\caption{Out-of-control ARL achieved for an additive shift in $\beta_{23}$ or $\beta_{24}$ by a factor of $c$ \label{tab:ARL1}} 
\begin{tabular}{@{}ccc@{}}
\hline
Change factor $c$ & Shift in & Shift in  \\ 
  & $\beta_{23}$  & $\beta_{24}$ \\
	\hline
			0.2 & 197.9 & 195.7 \\
			0.4 & 194.4  & 183.5 \\
			0.6 &190.2  & 169.5 \\
			0.8 &187.9  & 153.3 \\
			1.0 & 181.6 &  137.6\\
			1.2 & 174.8 & 123.8 \\
			1.4 & 166.8 & 112.8 \\
			1.6 & 157.9 & 102.4 \\
			1.8 &  148.4 & 95.1\\
			2.0 & 140.2 & 88.9 \\
			2.2 & 133.4 & 84.7 \\
			2.4 & 123.6 &  80.2\\
			2.6 & 117.4 &  77.1\\
			2.8 & 111.0 &  73.7\\
			3.0 & 102.8 &  72.2\\
			3.2 & 98.2 &  70.3\\
			3.4 & 91.4 & 69.1 \\
			3.6 & 86.9 & 68.3\\
			3.8 & 82.3  & 67.2 \\
			4.0 & 77.3 & 66.3 \\
			\hline
\end{tabular}
	\end{table}

In our next investigation we consider performance of the chart when there is a shift in the effect of an upstream outcome variable in a downstream outcome.  Such changes are likely to occur when changes in management of upstream adverse events occur that could be due to policy or staff changes. For illustration, changes of various sizes are introduced to $\gamma_{23}, \gamma_{24}$ and $\gamma_{34}$.  The results are shown in Table \ref{tab:ARL2} and show that detection of changes occurs more quickly for equivalent shifts in coefficients that are larger. In conjunction with the results in Table \ref{tab:ARL1}, we see that it is the size of the coefficient and size of shift that determines how quickly a change is detected.  The type of coefficient, that is whether it is associated with a process variable, upstream outcome or risk factor does not seem to have an impact on how quickly shifts are detected. This is to be expected since the risk factors and process variables play the same role in the model.  

\begin{table}[h]
\small\sf\centering
		\caption{Out-of-control ARL achieved for an additive shift in $\gamma_{23}$, $\gamma_{24}$ or $\gamma_{34}$  by a factor of $c$ \label{tab:ARL2}} 
\begin{tabular}{@{}cccc@{}} 
\hline
Change factor $c$ & Shift in & Shift in & Shift in\\
& $\gamma_{23}$ & $\gamma_{24}$ & $\gamma_{34}$ \\
\hline
0.2	& 198.3 &	197.6 &	196.8  \\
0.4	& 195.9	& 195.4 & 172.8	 \\
0.6	& 193.1	& 190.2 & 143.5	\\
0.8	& 193.8	& 179.2 & 129.4	\\
1.0 & 189.3	& 172.8 & 119.1	\\
1.2	& 187.5	& 155.8 & 88.4	\\
1.4	& 180.3	& 147.3 & 113.3 \\
1.6	& 173.6 & 137.0 & 62.8 \\
1.8	& 169.1	& 129.1 & 87.0 \\
2.0	& 163.7	& 114.5 & 22.0	\\
2.5	& 147.8 & 89.2  &	57.5 \\
3.0	& 129.7	& 78.1 & 97.5 \\
3.5	& 114.7	& 68.3 & 88.7	\\
4.0	& 100.0	& 53.4 & 60.0	\\
			\hline
\end{tabular}
	\end{table}

\begin{figure}[H]
\centering
\includegraphics[scale=0.6]{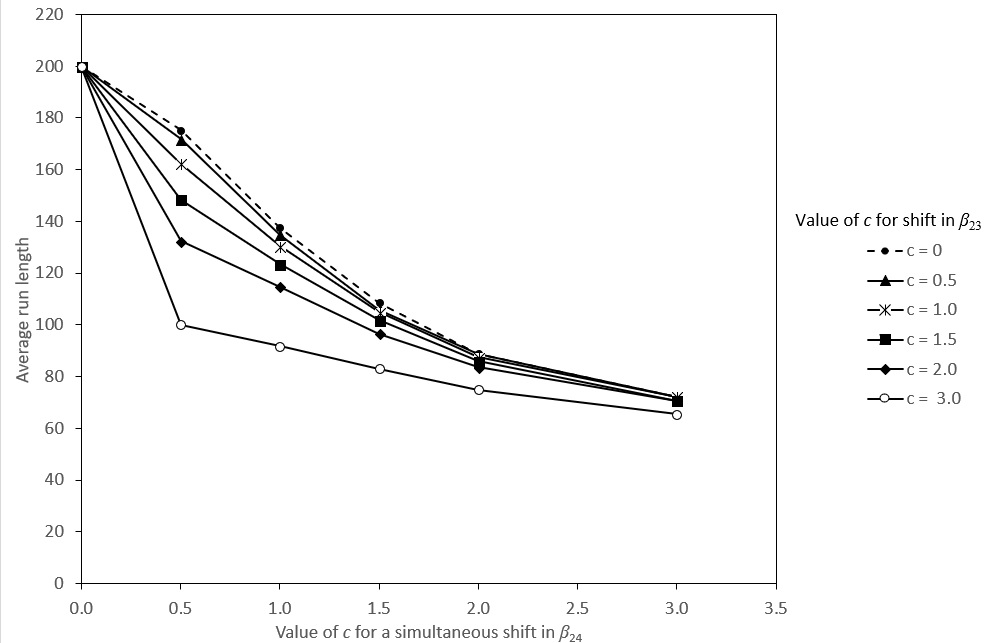}
\caption{\label{fig.2betas} Average run length for simultaneous shifts in $\beta_{23}$ and $\beta_{24}$. The dotted line shows the ARL for various shifts in $\beta_{24}$ when there is no concurrent shift in $\beta_23$.}
\end{figure}

\begin{figure}[H]
\centering
\includegraphics[scale=0.6]{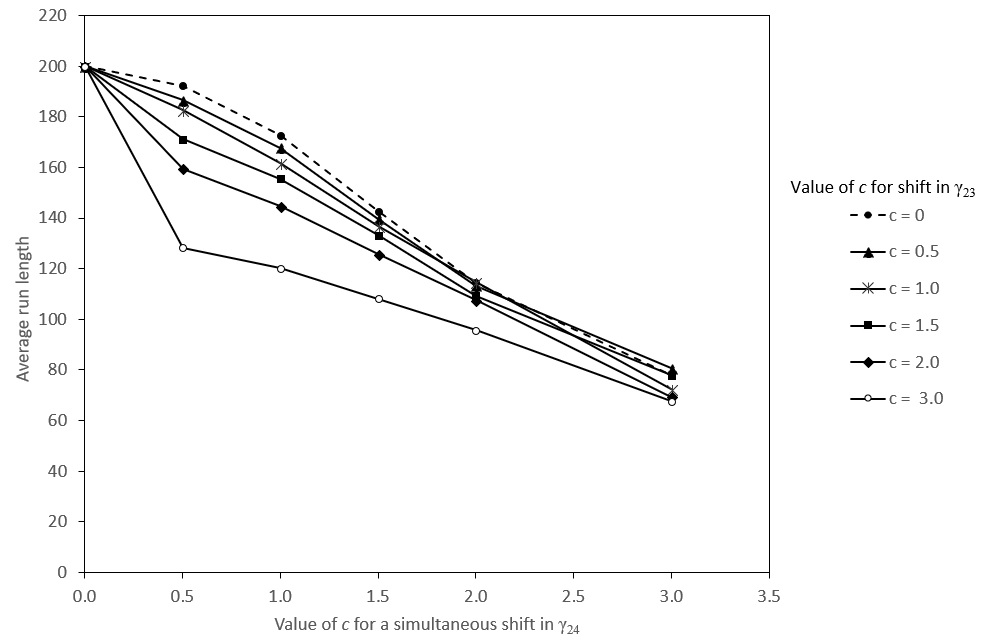}
\caption{\label{fig.2gammas} Average run length for simultaneous shifts in $\gamma_{23}$ and $\gamma_{24}$. The dotted line shows the ARL for various shifts in $\gamma_{24}$ when there is no concurrent shift in $\gamma_23$.}
\end{figure}

In our next investigation, we consider simultaneous shifts in more than one coefficient, indicating shifts in multiple parts on the process.  For multi-stage health care procedures where there is no tight control of the process unlike in manufacturing, it is realistic to expect that changes can occur simultaneously at various stages.  The chart we propose presents an approach for detection of any such changes at a global level.  For illustration, we consider simultaneous shifts in the pairs $(\beta_{23}, \beta_{24})$ and in $(\gamma_{23}, \gamma_{24})$. The results are shown in Figures \ref{fig.2betas} and \ref{fig.2gammas} and indicate that shifts in pairs of coefficients are detected more quickly than shifts in a single coefficient (dotted line).  

Next, we investigate chart performance when a shift occurs in the mean of an upstream outcome variable that affects a downstream outcome. Such a change represents an outcome shift whose source is unknown. A typical case where such a shift may occur is where small shifts occur in various parts of the functional relationship that collectively result in an overall mean shift.  These kinds of changes are likely to occur in most health procedures that are complex in nature with multiple contributing components. For illustration, we investigate performance of the chart when there is a shift in $\mu_{3t}$,  the probability of a 3rd or 4th degree tear occurring for patient $t$.  We consider both an additive shift and a shift in the odds ratio.  The results are shown in Table \ref{tab:ARL3}. As expected the larger the shift, the lower the out-of-control average run length.

	\begin{table}[h]
	\small\sf\centering
   \caption{ARL achieved for a shift in $E(Y_3)$ in the additive and odds ratio format by factor $c$ \label{tab:ARL3}} 
\begin{tabular}{@{}ccc@{}} 
\hline
Change factor $c$ & \multicolumn{2}{c}{ARL for:} \\ 
& Additive shift & Odds ratio shift \\
\hline
0.2 & 188.2 & 188.2 \\
0.4	& 169.7 & 138.1 \\                                                  
0.6	& 153.8 & 137.3 \\
0.8	& 138.5 & 135.3 \\
1.0 & 124.9 &	130.6 \\
1.2	& 114.0 & 123.2 \\
1.4	& 103.8 & 116.0 \\
1.6 & 95.1 &	109.5 \\
1.8	& 88.3 &  102.6 \\
2.0	& 82.0 & 96.5 \\
2.2	& 75.5 & 91.1 \\
2.4 & 71.1 & 86.4 \\
2.6	& 67.7 & 81.8 \\
2.8 & 63.8 & 78.1 \\
3.0	& 60.0 & 74.6 \\
3.2	& 57.1 & 71.3 \\
3.4	& 54.3 & 68.6 \\
3.6	& 52.3 & 65.7 \\
3.8	& 49.8 & 63.2 \\
4.0	& 46.9 & 61.1 \\
\hline
\end{tabular}
 \end{table}

Therefore the chart we propose has the ability to detect changes of various types in the process. The greater the shift, whether it is in a single coefficient, an outcome mean or in multiple coefficients, the quicker it is to detect.  The chart can therefore be used as a tool for on-going monitoring of a multi-stage procedure at the global level.  An out-of-control signal would then be followed by detailed investigations of various parts of the process.

\section{Discussion}
\label{sec.discussion}
In this article, we proposed a new multi-stage multivariate control chart based on likelihood score equations to monitor the outcomes of health care procedures. A multi-stage approach is used to track outcomes at all stages through monitoring the coefficients of a model that represents the relationships among the outcomes, processes and patient risk factors relevant to the procedure. To ensure quick detection of small gradual shifts we used a sequential monitoring approach where the chart statistic is updated at each successive observation. We use likelihood score equations to construct a multivariate EWMA chart statistic.  The advantage of our score-based MEWMA  proposed chart  is that it can be designed to detect small or large shifts. Moreover, our approach is flexible in that it can handle processes with different types of variables  since the likelihood function and score
equations can be obtained in a straightforward manner once the assumed distributions of
the response variables are known.
Using simulations, we demonstrated the sensitivity of our proposed approach in reacting to various shift in the process and  the effect of upstream outcomes on monitoring end stage outcomes. The chart showed efficient performance in detecting small shift sizes.  Inclusion of risk factors in the model mean that factors outside of the practitioner's control
are accounted for. Signals that occur should be followed up with a review of the cases with
adverse outcomes to determine the appropriate action to take.

%\section{Software}
%\label{sec5}
%
%Software in the form of R code, together with a sample
%input data set and complete documentation is available on
%request from the corresponding author (nokuthaba.sibanda@vuw.ac.nz).

%\section{Supplementary Material}
%\label{sec6}
%
%Supplementary material is available online at
%% \href{http://biostatistics.oxfordjournals.org}%
%% {http://biostatistics.oxfordjournals.org}.
%\url{http://biostatistics.oxfordjournals.org}.

\vspace*{\baselineskip}

{\it Conflict of Interest}: None declared.

\section*{Acknowledgements}
We thank Professers Stefan Steiner and Jock Mackay for useful discussions at the initial phase of this work.

\bibliographystyle{apalike}
\bibliography{multistage_OE2}

\end{document}